\documentclass[12pt]{article}
\usepackage{epsfig}

\newskip\humongous \humongous=0pt plus 1000pt minus 1000pt

\newif\ifdtup

\def\abs#1{\left| #1\right|}
\def\pr#1{#1^\prime}

\def\beq{\begin{equation}}
\def\eeq{\end{equation}}

\def\beqn{\begin{eqnarray}}
\def\eeqn{\end{eqnarray}}
\relax

\def\s{\sigma}
\def\fix{\right|_{\rm FO}\!\!\!\!}

\def\lq{\left[}
\def\rq{\right]}

\def\({\left(}
\def\){\right)}
\def\xe{x_E}
\def\xp{x_p}
\def\zp{z_p}

\catcode`\@=11

\@addtoreset{equation}{section}
\def\theequation{\thesection.\arabic{equation}}

\def\@normalsize{\@setsize\normalsize{15pt}\xiipt\@xiipt
\abovedisplayskip 14pt plus3pt minus3pt%
\belowdisplayskip \abovedisplayskip
\abovedisplayshortskip \z@ plus3pt%
\belowdisplayshortskip 7pt plus3.5pt minus0pt}

\def\small{\@setsize\small{13.6pt}\xipt\@xipt
\abovedisplayskip 13pt plus3pt minus3pt%
\belowdisplayskip \abovedisplayskip
\abovedisplayshortskip \z@ plus3pt%
\belowdisplayshortskip 7pt plus3.5pt minus0pt
\def\@listi{\parsep 4.5pt plus 2pt minus 1pt
     \itemsep \parsep
     \topsep 9pt plus 3pt minus 3pt}}

\@twosidetrue

\relax

\catcode`@=12

\catcode`\@=11

\def\section{\@startsection{section}{1}{\z@}{3.5ex plus 1ex minus
   .2ex}{2.3ex plus .2ex}{\large\bf}}

\def\thesection{\arabic{section}}

\def\appendix{\setcounter{section}{0}
 \def\thesection{\Alph{section}}
 \def\theequation{\Alph{section}.\arabic{equation}}
 \def\section{\@startsection{section}{1}{\z@}{3.5ex plus 1ex minus
   .2ex}{2.3ex plus .2ex}{\large\bf}}
 \def\subsection{\@startsection{section}{2}{\z@}{3.25ex plus 1ex minus
   .2ex}{1.5ex plus .2ex}{\large\bf}}}

\def\ps@headings{\def\@oddfoot{}\def\@evenfoot{}
\def\@oddhead{\hbox{}\hfill
 \makebox[.5\textwidth]{\raggedright\ignorespaces --\thepage{}--
 \hfill {}}}  
\def\@evenhead{\@oddhead}
\def\subsectionmark##1{\markboth{##1}{}}
}

\ps@headings

\catcode`\@=12

\def\figcap{\section*{Figure Captions\markboth
 {FIGURECAPTIONS}{FIGURECAPTIONS}}\list
 {Fig. \arabic{enumi}:\hfill}{\settowidth\labelwidth{Fig. 999:}
 \leftmargin\labelwidth
 \advance\leftmargin\labelsep\usecounter{enumi}}}
 \relax
\def\tablecap{\section*{Table Captions\markboth
 {TABLECAPTIONS}{TABLECAPTIONS}}\list
 {Table \arabic{enumi}:\hfill}{\settowidth\labelwidth{Table 999:}
 \leftmargin\labelwidth
 \advance\leftmargin\labelsep\usecounter{enumi}}}
 \relax
\def\reflist{\section*{References\markboth
 {REFLIST}{REFLIST}}\list
 {[\arabic{enumi}]\hfill}{\settowidth\labelwidth{[999]}
 \leftmargin\labelwidth
 \advance\leftmargin\labelsep\usecounter{enumi}}}
 \relax

\catcode`\@=11

\relax

\catcode `@ 11
\def\biblabel#1{\if@filesw\immediate
\write\@auxout{\string\bibcite{#1}{\the\value{\@listctr }}}\fi}
\catcode `@ 12

\def    \be             {\begin{equation}}
\def    \ee             {\end{equation}}
\def    \ba             {\begin{eqnarray}}
\def    \ea             {\end{eqnarray}}

\def    \=              {\;=\;}
\def    \frac           #1#2{{#1 \over #2}}

\def \ep{\epsilon}
\def \as   {\ifmmode \alpha_s \else $\alpha_s$ \fi}

\def\b0{b_0}

\def \mt   {\ifmmode m_{\rm t} \else $m_{\rm t}$ \fi}

\def \to   {\mbox{$\rightarrow$}}
\newcommand     \MSB            {\ifmmode {\overline{\rm MS}} \else
                                 $\overline{\rm MS}$\fi}

\newcommand\hepph[1]{{\tt hep-ph/#1}}

\newcommand     \epem           {\ifmmode{e^+e^-}\else{$e^+e^-$}\fi}
\def \to   {\mbox{$\rightarrow$}}

\def\figura#1#2#3
{\begin{figure}[hb]
  \begin{center}
  \leavevmode\protect\epsfxsize=#1\protect\epsffile{#2.eps}
  \protect\caption{#3}
  \label{#2}
  \end{center}
  \end{figure}
}
\newlength{\Largfig}
\def\({\left(}
\def\){\right)}

\def\s{\sigma}

\def\asb{{}\ifmmode \bar{\alpha}_s \else $\bar{\alpha}_s$\fi}

\def\OPAL{\mbox{OPAL}}

\def\ARGUS{\mbox{ARGUS}}

\def\ord#1{{\cal O}(#1)}

\relax
\def\pl#1#2#3{{\it Phys. Lett. }{\bf #1}\ (19#2)\ #3}

\def\prl#1#2#3{{\it Phys. Rev. Lett. }{\bf #1}\ (19#2)\ #3}
\def\rmp#1#2#3{{\it Rev. Mod. Phys. }{\bf#1}\ (19#2)\ #3}
\def\prep#1#2#3{{\it Phys. Rep. }{\bf #1}\ (19#2)\ #3}
\def\pr#1#2#3{{\it Phys. Rev. }{\bf #1}\ (19#2)\ #3}
\def\np#1#2#3{{\it Nucl. Phys. }{\bf #1}\ (19#2)\ #3}
\def\sjnp#1#2#3{{\it Sov. J. Nucl. Phys. }{\bf #1}\ (19#2)\ #3}
\def\app#1#2#3{{\it Acta Phys. Polon. }{\bf #1}\ (19#2)\ #3}
\def\jmp#1#2#3{{\it J. Math. Phys. }{\bf #1}\ (19#2)\ #3}
\def\nc#1#2#3{{\it Nuovo Cim. }{\bf #1}\ (19#2)\ #3}

\relax

\newcount\minutes
\newcount\scratch

\def\timestamp{%
\scratch=\time
\divide\scratch by 60
\edef\hours{\the\scratch}
\multiply\scratch by 60
\minutes=\time
\advance\minutes by -\scratch
---$\,$\hours:\null
\ifnum\minutes< 10 0\fi
\the\minutes}

\begin{document}
\begin{titlepage}
\nopagebreak
{\flushright{
        \begin{minipage}{4cm}
        CERN-TH/98-339  \hfill \\
        IFUM 637/FT \hfill \\
        DTP/98/76\\
        hep-ph/9811206\hfill \\
        \end{minipage}        }

}
\vfill
\begin{center}

{\LARGE \bf \sc
\baselineskip 0.8cm \noindent
A Fixed-Order Calculation of\newline\noindent
the Heavy-Quark Fragmentation\newline\noindent
Function in $e^+e^-$ Collisions

}
\vskip 2cm
{\bf Paolo NASON\footnote{From November 1$^{\rm st}$: INFN, Sezione di Milano,
  Milan, Italy},}
\\
\vskip 0.1cm
{TH Division, CERN, Geneva, Switzerland} \\
\vskip .5cm
{\bf Carlo OLEARI}
\\
\vskip .1cm
{Department of Physics, University of Durham, Durham, England}
\end{center}
\nopagebreak
\vfill
\begin{abstract}
We use a recently completed ${\cal O}(\as^2)$ fixed-order calculation
of the heavy-flavour production cross section in $e^+e^-$
collisions to compute the heavy-quark fragmentation function.
We fit the result of our calculation, convoluted with a Peterson
fragmentation function, to available data for charm production, and thus
obtain a value for the parameter $\ep$ in the Peterson function.
We discuss the relevance of mass effects and of subleading terms
in our calculation.
\end{abstract}
\vskip 1cm
CERN-TH/98-339 \hfill \\
October 1998
\vfill
\end{titlepage}

\section{Introduction}

In this work, we use a recently completed calculation of the
$\ord{\as^2}$ differential cross section for heavy-quark production in
$e^+e^-$ annihilation~\cite{zbb4}
to compute the heavy-quark fragmentation function at order
$\as^2$. This calculation should be reliable when the
centre-of-mass energy $E$ is not too high. At very high energies,
in fact, large logarithms of the ratio $E/m$, where $m$ is the 
heavy-quark mass, arise at all orders in perturbation theory,
and should be resummed. A method for the resummation of the large logarithms
at the next-to-leading logarithmic level (NLL) has been developed
in Ref.~\cite{MeleNason}.
On the other hand, the fixed-order calculation should be more
accurate for moderate values of the energy,
since it correctly accounts for mass effects.
Furthermore, the NLL calculation correctly accounts for terms
proportional to $\as^2\log^2(E/m)$ and $\as^2\log(E/m)$,
but cannot correctly predict the $\as^2$ terms that do not carry
any logarithmic enhancements,
since these terms are
of next-to-next-to-leading logarithmic order (NNLL).

Studies of the charm fragmentation function have been performed
with relatively recent data in Ref.~\cite{CacGre},
using a parametrization of the non-perturbative effects based upon the
Peterson fragmentation function. From these studies, 
it was found that the value of the $\ep$ parameter is much smaller
in NLL fits rather than in leading-log (LL) ones.
In this work, we  fit the same data sets, using our 
fixed-order calculation convoluted with a Peterson
parametrization of non-perturbative effects, and compare our results
with those of Ref.~\cite{CacGre}.
In order to better understand the differences of the two approaches,
we will also consider a fixed-order calculation of the fragmentation function,
in which mass-suppressed effects (i.e.\ effects suppressed by powers of
$m/E$) and NNLL terms
are neglected. This calculation corresponds to a truncation
of the NLL formalism at order $\as^2$.
\section{Theoretical framework}
We consider the inclusive production of a heavy quark
$Q$ of mass $m$
\beq 
\label{eq:process}
  e^+ e^- \,\to\, Z/\gamma\;(q) \,\to\, Q\,(p) + X\;,
\eeq
where $q$ and
$p$ are the four-momenta of the intermediate boson and of the final quark.
We also introduce the notation
\begin{equation}
 E=\sqrt{q^2}\,,\quad\quad \rho=\frac{4m^2}{q^2}\,.
\end{equation}
We will consider two possible definitions of the $x$ variable,
one based upon the energy and one based upon the momentum.
In the centre-of-mass system, we define
\beq
   \xe = \frac{p^0}{p^0_{\rm max}}\,,\;\quad\quad
   \xp = \frac{\abs{\vec{p}}}{\abs{\vec{p}_{\rm max}}}\,,
\eeq
with the kinematic ranges
\begin{equation}
  \sqrt{\rho}\le \xe \le 1 \,,\quad\quad 0\le \xp \le 1\,.
\end{equation}
In terms of invariants, we have
\begin{equation}
  \xe=\frac{2\,p\cdot q}{q^2}\,,\quad\quad
  \xp=\frac{\sqrt{\xe^2 - \rho}}{\sqrt{1-\rho}}\,.
\end{equation}
Our starting formula will be the fixed-order (FO)
cross section for the inclusive
production of a heavy-flavoured hadron.
It is given by the
convolution of the cross section for the inclusive production
of a heavy quark, supplemented
with a non-perturbative fragmentation function, which describes
phenomenologically all the large time phenomena related to the
hadronization process
\beqn
\label{eq:sig_fix_had}
 \left. \frac{d\s^H}{d\xp}(\xp,E,m)\fix &=& \int_0^1 dy\,d\zp
 \left. \frac{d\s}{d\zp}(\zp,E,m)\fix  P(y,\ep) \,\delta(\xp - y\zp)\;,
\eeqn
where $P(y,\ep)$ is the Peterson~\cite{Peterson} fragmentation function
\begin{equation}
  P(y,\ep)\equiv  N \,\frac{y\,(1-y)^2}{\lq (1-y)^2+y\,\ep\rq^2} \;,
\end{equation}
where the normalization factor $N$ is fixed by the condition
\beq
 \int_0^1  dy \, P(y,\ep) = 1\;
\eeq
if $P$ refers to the total fragmentation function (i.e., summed
over all heavy-flavoured hadron species). In the following,
where we will mostly consider $D^*$ production, the normalization
will be fitted to the data.
Notice that we have written the convolution in terms of the momentum
fraction, rather than the energy fraction. At large momenta, the
difference between the two definitions is small.
At small momenta, one could choose either approach.
Choosing $\xp$ seems, however, simpler and more sensible, since it
is more conceivable that at small momenta the non-perturbative effects soften
the hadron momentum, rather than its mass.

The details of the procedure we followed
to perform the calculation will be given in a forthcoming publication.
It is, however, quite clear that the heavy-quark inclusive cross section
$d\s/d\zp$ can be computed using the results of
Refs.~\cite{NasonWebber,zbb4}. In order to compute
the truncated NLL cross section, we have used the results of
Ref.~\cite{frag97}, where the NLL evolution equations,
with appropriate initial conditions, have been solved exactly up to
the second order in the strong coupling constant.

In the present calculation, we have neglected all contributions
to the heavy-flavour cross section arising from gluon splitting.
These contributions are small at moderate energies, and in general
affect the heavy-flavour inclusive cross section at small values of $x$.
\section{Phenomenological results}
\begin{figure}[htb]
\centerline{\epsfig{figure=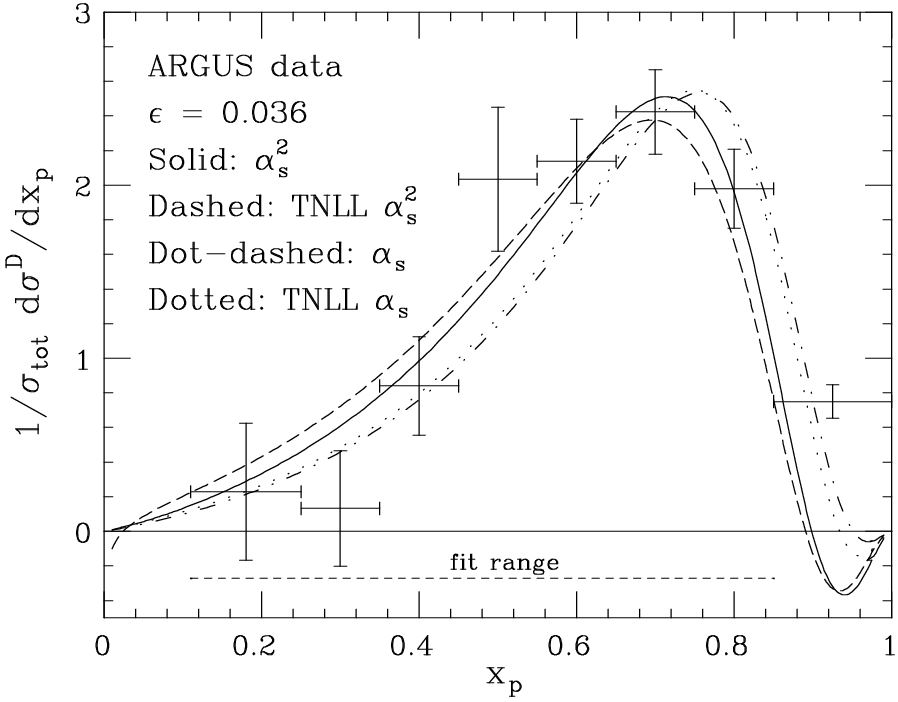,width=0.65\textwidth,clip=}}
\caption{ \label{fig:argus_fit}
Best fit for the ${\cal O}(\as^2)$ fragmentation function 
at {\ARGUS}.}
\end{figure}

Our main results are summarized in Fig.~\ref{fig:argus_fit},
where the fitted, ${\cal O}(\as^2)$ fragmentation function is shown
together with the ARGUS data for $D^{*+}$ production~\cite{ARGUS}.
The free parameters in the fixed-order calculation are the
charm-quark mass, which we have fixed at 1.5~GeV,
$\Lambda_{\rm QCD}^{(5)}$, which we have fixed to 200~MeV
(corresponding to $\as(M_Z)=0.116$)
$\epsilon$, which we have fitted, and the normalization,
which we have also fitted.
The large-$x$ bin has been excluded from the fit.
This is justified, since large logarithms of $(1-x)$ spoil the
accuracy of the perturbative expansion in that region. A manifestation of this
pathology can be seen in the computed fragmentation function, which becomes
negative at large $x$.
The result of the fit is $\ep=0.036$, with $\chi^2/{\rm dof}=0.853$.
In Fig.~\ref{fig:argus_fit} we also display
the ${\cal O}(\as)$ fixed-order result and the truncated expansion
of the NLL result (TNLL), both at orders $\as$ and $\as^2$.
All these curves are obtained with the same value of $\ep=0.036$.
We see, first of all, that the ${\cal O}(\as)$ fixed-order
result is harder than the ${\cal O}(\as^2)$ one.
In fact, if we attempt to fit the data using the ${\cal O}(\as)$ fixed-order
result, we obtain $\ep=0.058$, with $\chi^2/{\rm dof}=0.852$.
The TNLL, ${\cal O}(\as)$ result differs from the full ${\cal O}(\as)$
one only by terms that are suppressed by powers of the mass over the energy.
The curves in the figure seem to indicate that these effects are already quite
small for charm at ARGUS energy. The TNLL, ${\cal O}(\as^2)$ result
differs from the full ${\cal O}(\as^2)$ one by terms that are suppressed 
by powers of the mass over the energy, and by terms of order $\as^2$ which are
not multiplied by large logarithms of the mass over the energy (NNLL terms).
The figure suggests that the presence of these terms makes the
fragmentation function harder. Thus, a smaller value of $\ep$ would be obtained
if we fitted the data using the TNLL ${\cal O}(\as^2)$ result.

\begin{figure}[htb]
\centerline{\epsfig{figure=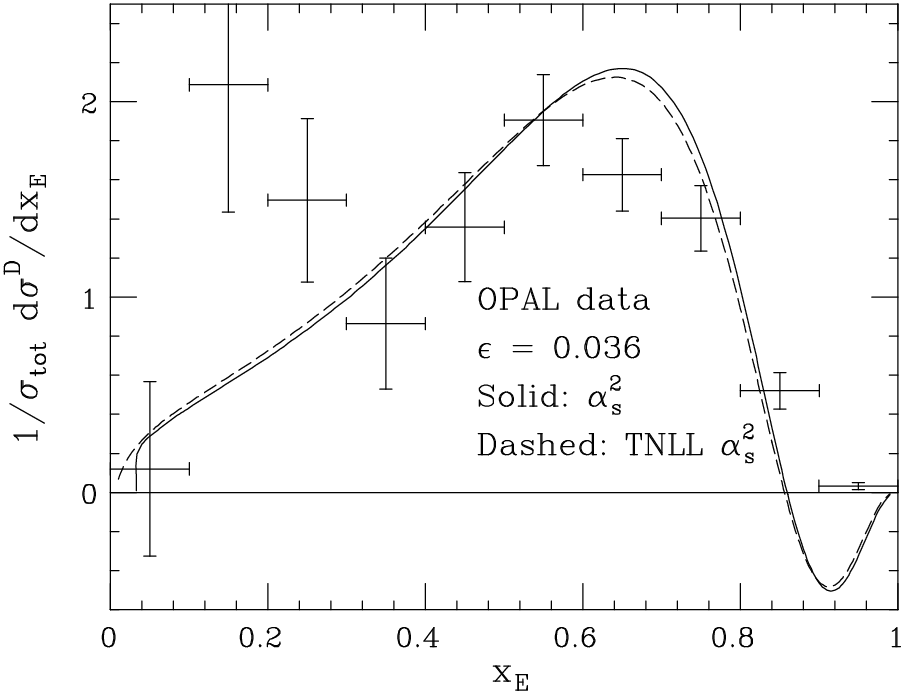,width=0.65\textwidth,clip=}}
\caption{ \label{fig:opal_036}
The ${\cal O}(\as^2)$ fragmentation function plotted together
with {\OPAL} data.}
\end{figure}

In Fig.~\ref{fig:opal_036} we plot the computed fragmentation function
at LEP1 energy, using the same value of $\ep=0.036$, together
with data from OPAL \cite{OPAL}.
The \OPAL\ data are in terms of $\xe$, and
we have thus performed the appropriate change of variable in
our cross section formulae. The data are arbitrarily normalized.
It is apparent from the figure that some evolution effect is present
in the fixed-order computation, so that the fragmentation function
is softer at higher energy. However, it is not quite as soft as the
data would require. If we fit the value of $\ep$ to the OPAL data,
we get $\ep=0.041$, a somewhat larger value than in the ARGUS case.
In this fit, besides excluding the large-$x$ region, we should also exclude
the small-$x$ bins, since our calculation does not include gluon
splitting effects, and these become more significant at high energy.
We also plot the TNLL, ${\cal O}(\as^2)$ result. We see that at this
energy it differs very little from the fixed-order result.
It is nevertheless difficult to disentangle mass effects from the
NNLL, ${\cal O}(\as^2)$ terms. In fact,
the former should be reduced
by a factor of 10 when going from ARGUS to LEP energies (assuming
a linear power law),
while the latter should be reduced (roughly) by a factor of 2,
due to the running in $\as^2$.
The figure seems to indicate something intermediate between these two values.

We now comment on the differences of our results with those of
Cacciari and Greco~\cite{CacGre}. These authors have fitted
the ARGUS data
using a resummed NLL calculation, and found the value $\ep=0.02$.
This value is considerably smaller than the commonly used value of 
$0.06$~\cite{Chrin}, which seems in fact to be appropriate only with
leading logarithmic calculations.
Our result confirms the fact that, when next-to-leading corrections
are introduced, smaller values of $\ep$ are needed.
On the other hand, our value of $\ep$ is larger. This is partly explained
by the comparison of our result with OPAL data. We expect that our
result will become worse as the energy increases,
and conversely, becomes better at lower energies.
Since our value of $\ep$ increases at higher energies, we expect
that it could decrease at lower energies, and thus approach the result
of Cacciari and Greco.
On the other hand, we have evidence that mass effects do make the
fragmentation function harder, and thus require a larger value of $\ep$
to fit the data.
To state this in a few words, we can say that our result tends to give
larger values of $\ep$ because it lacks resummation of leading and next-to-leading logarithms beyond the ${\cal O}(\as^2)$, while the result
of~\cite{CacGre} tends to give smaller values of $\ep$ because it lacks
mass effects.
\subsubsection*{Acknowledgements}
We wish to thank M.~Cacciari and C.~Grab for useful discussions.

\relax
\def\pl#1#2#3{{\it Phys.\ Lett.\ }{\bf #1}\ (19#2)\ #3}
\def\zp#1#2#3{{\it Z.\ Phys.\ }{\bf #1}\ (19#2)\ #3}
\def\prl#1#2#3{{\it Phys.\ Rev.\ Lett.\ }{\bf #1}\ (19#2)\ #3}
\def\rmp#1#2#3{{\it Rev.\ Mod.\ Phys.\ }{\bf#1}\ (19#2)\ #3}
\def\prep#1#2#3{{\it Phys.\ Rep.\ }{\bf #1}\ (19#2)\ #3}
\def\pr#1#2#3{{\it Phys.\ Rev.\ }{\bf #1}\ (19#2)\ #3}
\def\np#1#2#3{{\it Nucl.\ Phys.\ }{\bf #1}\ (19#2)\ #3}
\def\sjnp#1#2#3{{\it Sov.\ J.\ Nucl.\ Phys.\ }{\bf #1}\ (19#2)\ #3}
\def\app#1#2#3{{\it Acta Phys.\ Polon.\ }{\bf #1}\ (19#2)\ #3}
\def\jmp#1#2#3{{\it J.\ Math.\ Phys.\ }{\bf #1}\ (19#2)\ #3}
\def\nc#1#2#3{{\it Nuovo Cim.\ }{\bf #1}\ (19#2)\ #3}
\def\jhep#1#2#3{{\it J.\ High Energy Phys.\ }{\bf #1}\ (19#2)\ #3}
\relax

\end{document}